\documentclass[sigconf]{acmart}
\usepackage{multirow}
\usepackage{subcaption}
\usepackage{caption}
\usepackage{setspace}
\usepackage{amsmath}
\usepackage[english]{babel}
\definecolor{mygray}{gray}{0.5}
\usepackage{graphicx}
\graphicspath{{.} }
\usepackage{float}
\usepackage{flushend}
\usepackage{mathtools}
\usepackage{enumitem}

\definecolor{midnightgreen}{rgb}{0.0, 0.29, 0.33}

\copyrightyear{2020}
\acmYear{2020}
\setcopyright{acmcopyright}
\acmConference[SIGIR '20]{Proceedings of the 43rd International ACM SIGIR Conference on Research and Development in Information Retrieval}{July 25--30, 2020}{Virtual Event, China}
\acmBooktitle{Proceedings of the 43rd International ACM SIGIR Conference on Research and Development in Information Retrieval (SIGIR '20), July 25--30, 2020, Virtual Event, China}
\acmPrice{15.00}
\acmDOI{10.1145/3397271.3401323}
\acmISBN{978-1-4503-8016-4/20/07}

\begin{document}
\fancyhead{}

\title{Few-Shot Generative Conversational Query Rewriting}

\author{Shi Yu$^{*1}$, Jiahua Liu$^{*1}$, Jingqin Yang$^1$, Chenyan Xiong$^2$,}
\author{Paul Bennett$^2$, Jianfeng Gao$^2$, and Zhiyuan Liu$^1$}\authornote{The first two authors contributed equally.}
\affiliation{Tsinghua University$^1$, Microsoft Research AI$^2$ 
} 
\affiliation{
\texttt{\{yus17, yang-jq17\}@mails.tsinghua.edu.cn};
\texttt{alphaf52@gmail.com};
}
\affiliation{
\texttt{\{chenyan.xiong, Paul.N.Bennett, jfgao\}@microsoft.com};
\texttt{liuzy@tsinghua.edu.cn}
}

\renewcommand{\shortauthors}{Yu and Liu, et al.}

\begin{abstract}
Conversational query rewriting aims to reformulate a concise conversational query to a fully specified, context-independent query that can be effectively handled by existing information retrieval systems.
This paper presents a few-shot generative approach to conversational query rewriting. We develop two methods, based on rules and self-supervised learning, to generate weak supervision data using large amounts of ad hoc search sessions, and to fine-tune GPT-2 to rewrite conversational queries.  
On the TREC Conversational Assistance Track, our weakly supervised GPT-2 rewriter improves the state-of-the-art ranking accuracy by 12\%, only using very limited amounts of manual query rewrites. 
In the zero-shot learning setting, the rewriter still gives a comparable result to previous state-of-the-art systems.
Our analyses reveal that GPT-2 effectively picks up the task syntax and learns to capture context dependencies, even for hard cases that involve group references and long-turn dependencies.
\end{abstract}





\maketitle

\section{Introduction}
\label{sess:intro}


Recent advances in deep learning and text understanding facilitate the transition of information retrieval systems from keyword-based queries and ``ten-blue'' links to more conversational experiences.
Widely viewed as a next generation IR direction, Conversational IR is favored with its ability to satisfy users' complex information needs with multi-round interactions, while also providing convenient and precise information access through conversational interfaces and portable devices.


\begin{table}
  \caption{A Conversational Search Example in TREC CAsT}
  \label{tab:motivation}
  \begin{tabular}{cl}
    \toprule
    \midrule
    \multicolumn{2}{l}{\textbf{Description}:}                                           \\
    \multicolumn{2}{l}{The Bronze Age collapse and the transition into a dark age.}     \\
    \midrule
    \textbf{Turn}   & \textbf{Conversational Queries}                                   \\
    \midrule
    $Q_1$           & Tell me about the Bronze Age collapse.                            \\
    $Q_2$           & What is the evidence for it?                                      \\
    $Q_3$           & What are some of the possible causes?                             \\ 
    \midrule
    \multicolumn{2}{l}{\textbf{Manual Query Rewrites}}                                  \\ 
    \midrule
    $Q^*_2$         & What is the evidence for \textbf{the Bronze Age collapse}?        \\
    $Q^*_3$         & ... the possible causes \textbf{of the Bronze Age collapse}?      \\
    \midrule
    \bottomrule
  \end{tabular}
\end{table}

A signature of Conversational IR is its multi-round interactions with the user, an opportunity to understand and assist with more complex tasks and a challenge to query understanding. Natural conversations are concise and context dependent. Statements refer to previous discussions, omit already mentioned concepts, and assume implicit context during the conversation.
Table~\ref{tab:motivation} shows one such example from TREC Conversational Assistance Track (CAsT) 2019. The user begins with a fully specified query ($Q_1$), but quickly starts to use references ($Q_2$) and omissions ($Q_3$), which is very different from typical keyword-based search sessions.


A natural direction to tackle this challenge is to rewrite the conversational queries to de-contextualized queries that include all necessary information. The manually rewritten queries ($Q_2^*$ and $Q_3^*$ in Table~\ref{tab:motivation}) can be much better handled by existing ad-hoc ranking systems.
In TREC CAsT 2019, various approaches were developed for this \textit{conversational query rewriting} task, including IR-style query expansion/term reweighting, NLP-style coreference resolution, and neural-based query rewriting.
Still, conversational query rewriting is a challenging task: there is 30\%+ NDCG drop from systems that use automatic query rewriting/reformulation, compared with their counterparts using manual rewrites~\cite{jeffrey2019trec}.

One top performing conversational query rewriting system in TREC CAsT is ATeam's GPT-2 generative query rewriter (a later version can be found in \citep{Vakulenko2020QuestionRewriting}). They feed into a pre-trained transformer language model~\cite{radford2019language} the previous and current queries in the session (e.g. $Q_1, Q_2$ and $Q_3$), and fine-tune the model to generate the fully de-contextualized query rewrite ($Q_3^*$). 
The effectiveness and simplicity of this generative model make it a promising solution for conversational search. 
However, their GPT-2 was trained using their large quantity of manual query rewrites on their own conversational search queries.
It is not clear whether the transformer language model can still be effectively learned without large amounts of manual query rewrite labels, which are expensive to collect and are not always available for many domains~\cite{jeffrey2019trec}. 


This work studies learning with GPT-2 in conversational query rewriting using few or even zero manual rewriting labels.
We propose two approaches that generate weak supervision signals for this task using the ad hoc search sessions abundant in search logs.
The first is a rule-based approach which uses two simple rules to omit or co-refer repeated noun phrases in search sessions.
The second is a self-supervised learning approach that uses a handful of manually created query rewrites and conversational queries to train a GPT-2 model, as a simplifier, to convert the ad hoc search sessions to more context-dependent, conversational-like queries.
These approaches provide large amounts of weak supervision data for the GPT-2 rewriter to learn the context dependencies in concise conversational search queries.

In the few-shot setting where only TREC CAsT's manual query rewrites of 50 conversational sessions are used, ranking with our query rewrites outperforms the best automatic runs in CAsT 2019 by 12\% NDCG@3.
In the zero-shot setting, where no manual query rewrites are used, our weakly supervised GPT-2 still gives comparable result to the previous best automatic run in CAsT.

We further explore the capability of GPT-2 in few-shot learning by fine-tuning only on a handful of manual query rewrites. We observe that, surprisingly, the pre-trained transformer is able to pick up this task with \textit{as few as three conversational sessions}.
We find that GPT-2 quickly and effectively learns the task syntax: to generate questions instead of stories and to resolve the context dependencies using previous turns. We also observe that the model accurately deals with hard cases such as ones containing long-term and multiple coreferences.\footnote{Our code, data, and analyses results are publicly available at \url{https://github.com/thunlp/ConversationQueryRewriter}.}



\section{Preliminaries}
\label{sess:model}
This section describes the application of GPT-2 on the conversational query rewriting task.


\textbf{Conversational Query Rewriting.}
Conversational search systems aim to find relevant documents for queries in a conversational search session $S=\{Q_1,...Q_k...,Q_N\}$~\cite{jeffrey2019trec}. 
The conversational queries are often concise and their information needs are often presented in the previous queries.

The conversational query rewriting task is to rewrite a context dependent query $Q_k$ to a fully de-contextualized query $Q'_k$, with the help of previous queries $Q_{<k}$: 
\begin{align}
    Q'_k &= \texttt{QueryRewriter}(Q_k; Q_{<k}),
\end{align}
which better reflects user intent and is easier for ad hoc search.

We use GPT-2~\cite{radford2019language, jeffrey2019trec} to directly generate the query words $\{w'_1,...$ \newline $w'_i...,w'_M\}$ in $Q'_k$ one by one as:
\begin{align}
    w'_i &= f(w'_{<i}; Q_{k}, Q_{<k}). \label{eq:nlg}
\end{align}
where $f$ is transformer decoder and the input is in the format of:
\begin{align}
    Q_1 \circ [SEP] \circ ... \circ [SEP] \circ Q_k \circ [BOS] \circ [w'_1,...,w'_{i-1}],
\end{align}


Both training and inference use standard GPT-2~\cite{radford2019language}, which is adapted to our task to generate queries instead of plain text~\cite{jeffrey2019trec}. In training, the target query $Q_k^*=\{w_1^*...w_m^*\}$, either ground truth labels or weak supervision labels, are used to train the model.

\textbf{Ranking with Query Rewrites.} With the de-contextualized query rewrite $Q'_k$, standard ad hoc ranking can be used to complete the conversational search task. We use the standard BM25 to retrieve 100 documents and a BERT ranker to rerank them~\cite{Nogueira2019PassageRW, devlin-etal-2019-bert}.
\section{Weak Supervision}
\label{sess:weak-supervision}
One concern of generative query rewriting is that gold query rewrites $Q^*$ are expensive to obtain. This section describes how we leverage the ad hoc search sessions, available in search logs, to construct weak supervision data to mimic conversational search sessions with target query rewrites.

As current search engines are still moving towards conversational experiences, a typical ad hoc session is less likely to include many coreferences or omissions. Users may not expect search engines to resolve context dependency and tend to write fully specified queries. These fully specified queries, on the other hand, can be used as $Q^*$ in the conversational query rewriting task.

We consider ad hoc search sessions as pseudo target query rewrites, $\Tilde{S^*} = \{\Tilde{Q}^*_1,...\Tilde{Q}^*_i...,\Tilde{Q}^*_N$\}, and convert them to conversation-like sessions: $\Tilde{S} = \{\Tilde{Q}_1,...\Tilde{Q}_i...,\Tilde{Q}_N\}$. Then ($\Tilde{S}$, $\Tilde{S}^*$) pairs can serve as weak supervision to approximate real conversational queries $S$ and manual query rewrites $S^*$. 

To perform this conversion ($\Tilde{S^*} \rightarrow \Tilde{S}$), we propose two approaches, based on rules and self learning, respectively.

\textbf{Rule-Based.} The first approach uses two simple rules to mimic two discourse phenomena in conversations: \textit{omission} and \textit{coreference}. We perform the following operations on search sessions:
\begin{itemize}
\item \texttt{Omission.} A noun phrase is omitted if it occurs after a preposition and appears in previous queries; 
\item \texttt{Coreference.} Otherwise, previously appeared singular and plural noun phrases are respectively replaced with "it" (96\%), "he" (2\%), or "she" (2\%), and "they" (75\%) or "them" (25\%).
\end{itemize}
Both operations can be done efficiently on a vast amount of sessions.

\textbf{Self-Learn.} The second approach uses self-supervised learning and trains a GPT-2 model, known as query simplifier, to generate the conversation-like sessions $\Tilde{S}$ using $\Tilde{S}^*$. Differing from query rewriting that aims to ``put contexts back'' to the query, the query simplifier learns to generate contextual queries containing few information presented in previous queries of the same session.

The query simplifier uses a handful manual query rewrites, and learns to simplify the fully specified query to a contextual query as:
\begin{align}
    Q_k &= \texttt{QuerySimplifier}(Q^*_k; Q_{<k}).
\end{align}
Except reversing the source and target ($S^* \rightarrow S$), the same GPT-2 setup described in the previous section is used. The query simplifier, trained with a few manual query rewrites, is then applied to the ad hoc search sessions (MS MARCO) to generate more conversation-like sessions ($\Tilde{S^*} \rightarrow \Tilde{S}$).

\section{Experimental Methodologies}
Our experiments use the TREC CAsT 2019 benchmark for evaluation and the ad hoc sessions from MS MARCO for weak supervision.

\begin{table}
  \caption{Overall Results on TREC CAsT 2019 Conversational Search Task. \texttt{*} marks scores from \cite{jeffrey2019trec}. All our runs use the same ranking model. BLEU-2 are compared with Oracle Queries. QA-ROUGE evaluates the answer quality.}
  \label{tab:main-result}
  \small
  \begin{tabular}{lccc}
    \hline \hline
    \textbf{Method}                             & \textbf{BLEU-2}   & \textbf{NDCG@3}   & \textbf{QA-ROUGE}    \\
    \hline
    \multicolumn{4}{l}{\textbf{TREC CAsT Auto Runs}}                                                        \\
    \texttt{clacBase*}                           & --               & 0.360             & --                \\
    \texttt{pgbert*}                             & --               & 0.413             & --                \\
    \texttt{CFDA\_CLIP\_RUN7*}                   & --               & \textbf{0.436}    & --                \\
    
    \midrule
    \multicolumn{4}{l}{\textbf{CAsT Queries}}                                                               \\
    \texttt{Original}                            & 0.659            & 0.304             & 0.231             \\
    \texttt{AllenNLP Coref w/o sw}               & --               & 0.314             & --                \\
    \texttt{AllenNLP Coref w/ sw}                & 0.750            & 0.437             & 0.278             \\
    \texttt{Oracle}                              & 1.000            & 0.544             & 0.314             \\
    \midrule
    \multicolumn{4}{l}{\textbf{Zero-Shot Rewriter}}                                                         \\ 
    \texttt{GPT-2 Raw}                           & 0.112            & 0.124             & 0.196             \\
    \texttt{MARCO Raw}                           & 0.380            & 0.172             & 0.183             \\
    \texttt{Rule-Based}                          & \textbf{0.755}   & \textbf{0.437}    & 0.266             \\
    \midrule
    \multicolumn{4}{l}{\textbf{Few-Shot Rewriter}}                                                          \\
    \texttt{Rule-Based + CV w/o PLM}             & 0.178            & 0.065             & 0.151             \\
    \texttt{Self-Learn}                          & 0.750            & 0.435             & 0.263             \\
    \texttt{CV}                                  & 0.793            & 0.467             & 0.280             \\
    \texttt{Rule-Based + CV}                     & \textbf{0.809}   & \textbf{0.492}    & \textbf{0.291}    \\
    \texttt{Self-Learn + CV}                     & 0.804            & 0.491             & \textbf{0.291}    \\
     \hline \hline
  \end{tabular}
\end{table}

\textbf{TREC CAsT Conversation Search Benchmark.}
The dataset consists of 50 conversational search sessions $S$, each containing around ten conversational queries. The task is to retrieve and rank relevant passages for each query in $S$ from the MS MARCO passage collection and TREC Complex Answer corpora. 
Standard TREC relevance judgments are provided.
CAsT provides official manually rewritten queries for 50 conversational topics ~\cite{jeffrey2019trec}. 
We also manually label answer text for TREC CAsT questions and evaluate question answering result.\footnote{The answers are available at \url{https://github.com/thunlp/ConversationQueryRewriter}.}

\textbf{Evaluation Metrics.} The main metric in CAsT is NDCG@3 averaged on all turns. We also evaluate the similarity between automatic rewrites and ground truth using BLEU-2 and the question answering result using ROUGE-L.

\textbf{Weak Supervision Dataset and Preprocessing.} The ad hoc search sessions are collected from MS MARCO\footnote{https://github.com/microsoft/MSMARCO-Conversational-Search}.
It includes 152K artificial sessions, with MS MARCO queries automatically aligned to Bing search sessions. We process the DEV sessions to contain more question-like queries, by only retaining those with question words, and converting them to the weak supervision data (Sec.~\ref{sess:weak-supervision}).

\textbf{Baselines.} We compare with the following query reformation baselines. They all use the same ad hoc ranking as ours.

\texttt{Original} uses the original queries from TREC CAsT.

\texttt{AllenNLP Coref} uses the query reformulations (with or without stopwords) provided by CAsT where AllenNLP is used to resolve coreferences in search sessions. 

\texttt{GPT-2 Raw} directly applies the pre-trained GPT-2 for query rewriting  without fine-tuning. 

\texttt{MARCO Raw} fine-tunes GPT-2 on MS MARCO sessions for a language modeling task instead of the rewriting task.

\texttt{Oracle} uses the ground truth query rewrites provided by CAsT. This is the oracle run and falls in the manual category of CAsT~\cite{jeffrey2019trec}.

We also include three automatic runs from CAsT: \texttt{clacBase}, an expert query reformulation system, \texttt{pgbert}, a GPT-2 rewriter with external manual labels, and \texttt{CFDA\_CLIP\_RUN7}, a BERT based query expansion system.
The last two systems achieve the highest ranking accuracy among all automatic runs in CAsT 2019~\cite{jeffrey2019trec}. 


\textbf{Implementation Details.} The query rewriter is initialized using the pre-trained GPT-2 (medium) in Pytorch-Transformers. 

In the \textit{zero-shot} setting, only the weak supervision data of the converted MARCO sessions are used to fine-tune GPT-2. We include for comparison two \texttt{Raw} baselines and our \texttt{Rule-Based} method.

In the \textit{few-shot} setting, we also fine-tune on manual rewrites via five-fold cross validation (CV). 
We split the folds by sessions and \textit{no testing fold is revealed to model training}. Our methods in this setting include \texttt{Rule-Based + CV w/o PLM)}, \texttt{Self-Learn}, \texttt{CV}, \texttt{Rule-Based + CV}, and \texttt{Self-Learn + CV}.
We refer readers to our code repo for details.

Our GPT-2 uses batch size 2, learning rate 5e-5, and max sequence length 150.
Fine-tuning on weak supervision data converges after one epoch. 
Cross validation runs until convergence.  

The ad hoc ranking uses Anserini BM25 with INQUERY stopword removal.
The BERT ranker fine tunes BERT (base) \textit{only} using MS MARCO passage ranking labels; the CAsT relevance labels are only used in testing; our results are directly comparable with CAsT runs.

\section{Evaluation Results}
This section evaluates the effectiveness of our query rewriter in conversational search and analyzes the behavior of GPT-2.

\begin{figure}[t]
    \centering
    \begin{subfigure}[t]{0.45\columnwidth}
        \centering
        \includegraphics[width=\linewidth]{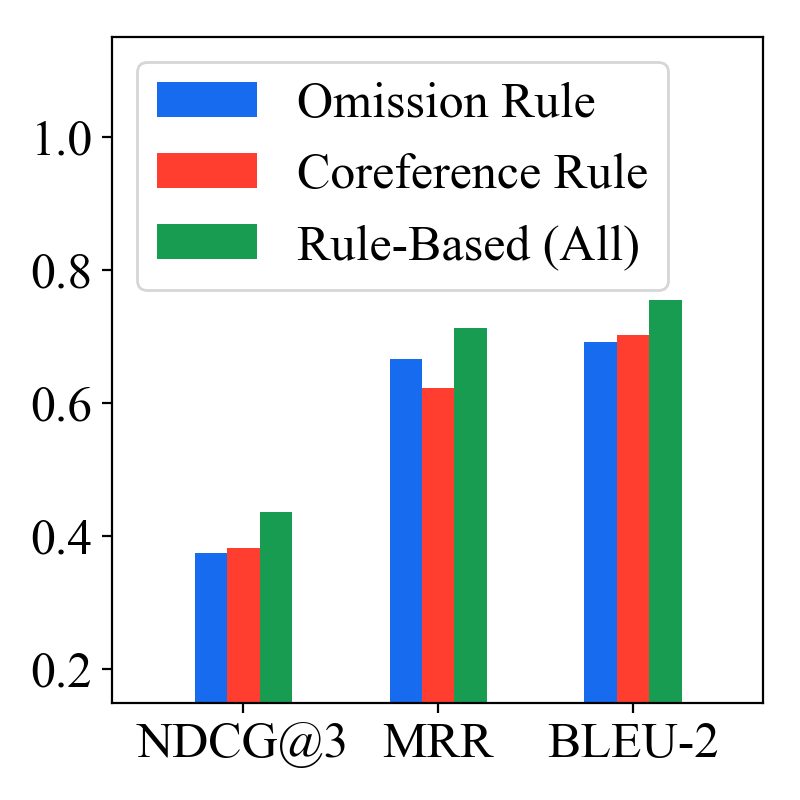}
        \caption{Different Rules}
        \label{fig:ablation-a}
    \end{subfigure}
    \begin{subfigure}[t]{0.45\columnwidth}
        \centering
        \includegraphics[width=\linewidth]{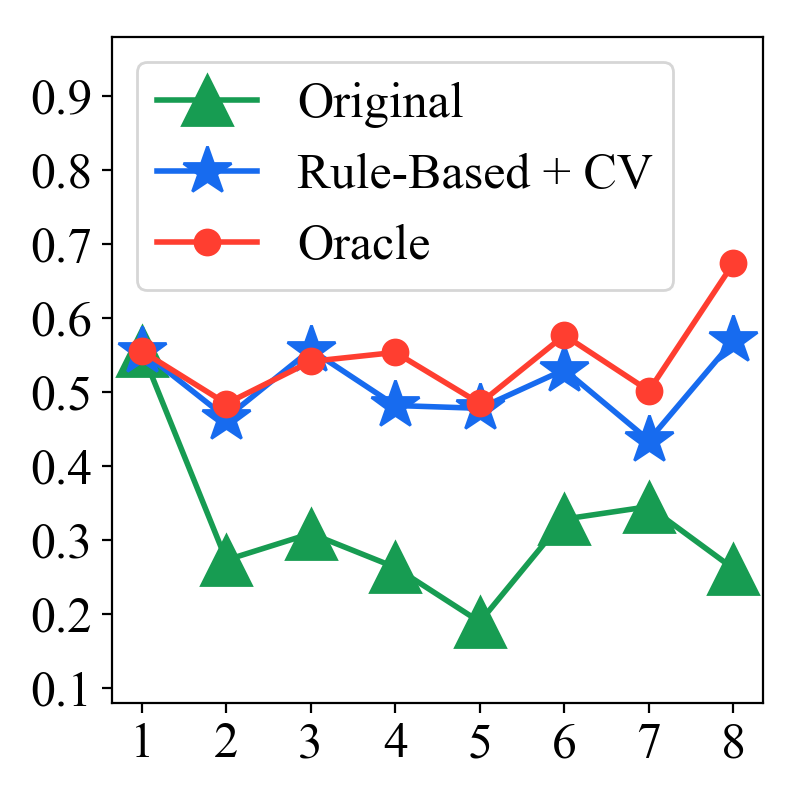}
        \caption{Conversational Depth}
        \label{fig:ablation-b}
    \end{subfigure}
    \caption{Performances in Different Scenarios. X-axis in (b) shows turn depths and Y-axis is NDCG@3.}
    \Description{}
\end{figure}

\subsection{Conversational Search Accuracy}
\label{subsess:main-result}

The overall Results in TREC CAsT are presented in Table~\ref{tab:main-result}.
As expected, the concise and contextual dependent nature of conversational search challenges existing ad hoc ranking and coreference resolution systems:
There is a significant gap between \texttt{Original} or \texttt{AllenNLP Coref} and manual \texttt{Oracle} queries. However, the gap is substantially narrowed by our GPT-2 query rewriter.

In the few-shot setting, GPT-2 trained with \texttt{CV} already outperforms the best CAsT auto runs, \texttt{pgbert} and \texttt{CFDA}.
Together with weak supervision data, \texttt{Rule-Based + CV} or \texttt{Self-Learn + CV} improves the state-of-the-art by 10+\%.   
The improvement is mainly attributed to better query rewriting: 
our simple BERT (base) ranker, when using \texttt{Oracle} queries, is less effective than \texttt{pgbert} and \texttt{CFDA} teams' manual runs; they obtained 0.57+ NDCG@3, compared to ours 0.544~\cite{jeffrey2019trec}.
The BLEU scores correlate well with NDCG---better query rewriting leads to better search accuracy.
Our query rewriter also maintains a stable accuracy in later turns, as shown in Fig.~\ref{fig:ablation-b},
which indicates that our rewriter effectively captures the multi-turn context as the conversation proceeds.

Surprisingly, GPT-2 (\texttt{CV}) provides effective rewrites when only cross validated on 50 CAsT sessions; \texttt{Rule-Based}, in the zero-shot setting,  is on par with best TREC CAsT automatic runs (Fig.~\ref{fig:ablation-a} shows their individual effectiveness).
In comparison, directly applying (\texttt{GPT-2 Raw}) or only fine-tuning using ad hoc sessions (\texttt{MARCO Raw}) yield sub-par results.
It is impressive that the pre-trained transformer can learn conversational query rewriting, a challenging task for previous techniques, in such a data efficient manner.

\subsection{Few-Shot Study}
This study further investigates GPT-2's capability of generalization.

\begin{figure}[t]
    \centering
    \begin{subfigure}[t]{0.45\columnwidth}
        \centering
        \includegraphics[width=\linewidth]{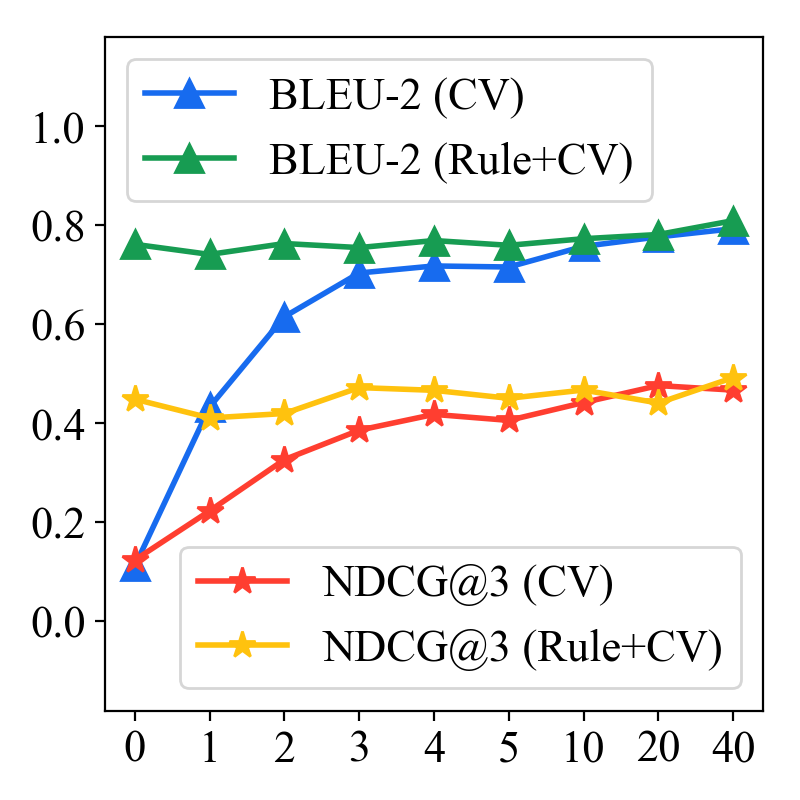}
        \caption{Training Sessions}
        \label{fig:analysis-a}
    \end{subfigure}
    \begin{subfigure}[t]{0.45\columnwidth}
        \centering
        \includegraphics[width=\linewidth]{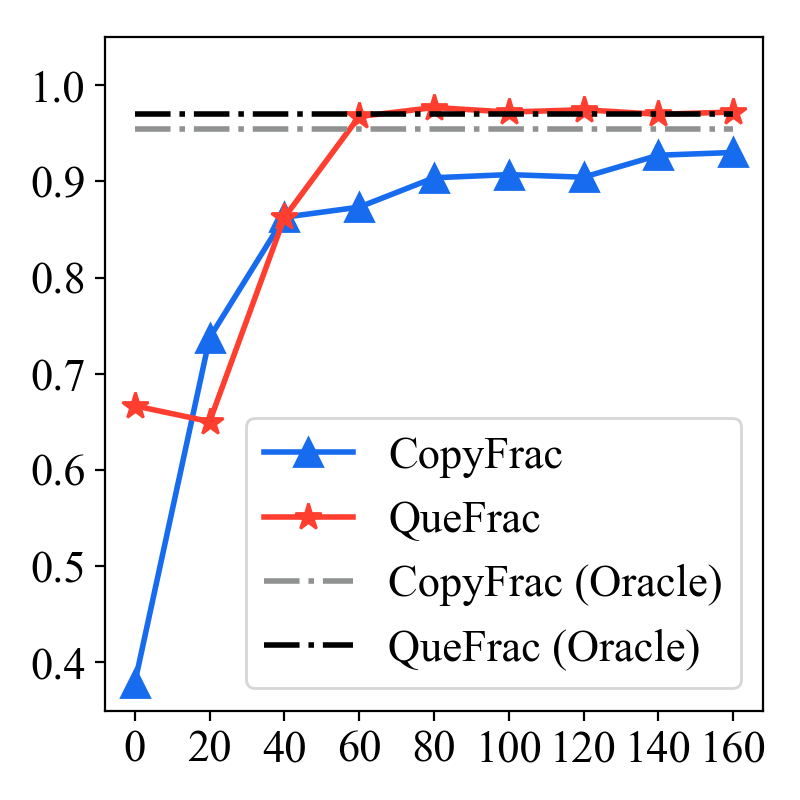}
        \caption{Training Steps}
        \label{fig:analysis-b}
    \end{subfigure}
    \caption{Performances of GPT-2 with different fine-tuning amounts: conversational sessions with manual rewrites (a) and fine-tuning steps (b). The Y-axes show the corresponding metric in (a) and (b).
    }
    \Description{}
\end{figure}

\textbf{How Few Shot?} 
Fig.~\ref{fig:analysis-a} shows GPT-2 fine-tuned with fewer sessions, with or without weak supervision.
Exceptionally, GPT-2 learns to generate reasonable query rewrites with \textit{only three conversational sessions or 30 manual labels}; it matches best CAsT auto runs with as few as 10 sessions.

\textbf{What is Learned?} 
It is unlikely that GPT-2 learns the discourse phenomena from just three sessions. 
They are likely to be captured in pre-training since the non-pre-trained \texttt{GPT-2} does not outperform substantially random guess, as in Table~\ref{tab:main-result}.

We hypothesize that GPT-2 only needs to learn the ``syntax'' of the rewriting task during fine-tuning: to generate questions and to replace pronouns with or add concepts mentioned in previous turns.
Fig.~\ref{fig:analysis-b} plots the fraction of questions (QueFrac) in  GPT-2 (\texttt{CV}) rewrites, indicated by question words, and the percentage of new words being copied from previous queries (CopyFrac), at different fine-tuning steps.
GPT-2 adapts to query rewriting very quickly with very little fine-tuning. Our effectiveness perhaps is more from properly ``unleashing'' the language understanding power already in the pretrained language model.

\subsection{Case Study}
Table~\ref{tab:case-study} provides two examples from GPT-2 (\texttt{Rule-based + CV}).
We found it surprising that in the first case, GPT-2 accurately resolves the group coreference from ``their'' to two cancer types, with one of the two from three turns ago. 
The second example presents a common error made by our rewriter: it fails to add proper context perhaps because in this case it is not clear what the context the term ``about'' refers to.
In our manual analyses, we found that GPT-2's errors are more often due to missing complete contexts than due to adding false information.



\begin{table}[t]
  \caption{GPT-2 Query Rewrites on CAsT Topic 31 and 64.}
  \label{tab:case-study}
  \begin{tabular}{cl}
    \hline \hline
    $Q_6$           & What causes \textbf{throat cancer}?                                   \\
    $Q_7$           & What is the first sign of it?                                         \\
    $Q_8$           & Is it the same as \textbf{esophageal cancer}?                         \\ 
    $Q_9$           & What's the difference in \underline{their} symptoms?                     \\
    \midrule
    \multirow{2}{*}{Oracle} & What's the difference in \textbf{throat cancer and}                   \\
                    & \textbf{esophageal cancer's} symptoms?                                \\
    \midrule
    \multirow{2}{*}{Output} & What's the difference between \textbf{throat cancer}                  \\
                    & \textbf{and esophageal cancer}?                                       \\
    \hline \hline
    $Q_1$           & What are the types of \textbf{pork ribs}?                             \\
    $Q_2$           & What are baby backs?                                                  \\
    $Q_3$           & What are the differences with spareribs?                              \\
    $Q_4$           & What are ways to \textbf{cook} them?                                  \\
    $Q_5$           & How \underline{about} on the bbq?                                                 \\
    \hline
    Oracle          & How \textbf{do you cook pork ribs} on the bbq?                        \\
    \hline
    Output          & How about on the bbq?                                                 \\
    \hline \hline
  \end{tabular}
\end{table}

\section{Conclusion}
This work demonstrates the effectiveness of GPT-2 for conversational query rewriting.
Fine-tuned using weak supervision data generated by rules or a handful of manual rewriting labels, our GPT-2 query rewriter is able to create new state-of-the-art on the TREC CAsT conversational search benchmark---outperforming previous methods including query expansion, contextual ranking, and coreference resolution, many of which use large-scale pre-trained models and deep neural networks.

\section*{Acknowledgements}
This work is supported by the National Key Research and Development Program of China (No. 2018YFB1004503) and the National Natural Science Foundation of China (NSFC No. 61732008, 61532010).

\bibliographystyle{ACM-Reference-Format}
\normalsize
\bibliography{citation}
\flushend 

\end{document}